[1]**Abildayeva T.,**
Master student, ORCID ID: 0009-0005-2983-3377
e-mail: t_abildayeva@kbtu.kz
[1]***Shamoi P.,**
PhD in Engineering, Professor, ORCID ID: 0000-0001-9682-0203,
*e-mail: p.shamoi@kbtu.kz

[1]Kazakh-British Technical University, 050000, Almaty, Kazakhstan

# FUZZY LOGIC APPROACH FOR VISUAL ANALYSIS OF WEBSITES WITH K-MEANS CLUSTERING-BASED COLOR EXTRACTION

**Abstract**

Websites form the foundation of the Internet, serving as platforms for disseminating information and accessing digital resources. They allow users to engage with a wide range of content and services, enhancing the Internet's utility for all. The aesthetics of a website play a crucial role in its overall effectiveness and can significantly impact user experience, engagement, and satisfaction. This paper examines the importance of website design aesthetics in enhancing user experience, given the increasing number of internet users worldwide. It emphasizes the significant impact of first impressions, often formed within 50 milliseconds, on users' perceptions of a website's appeal and usability. We introduce a novel method for measuring website aesthetics based on color harmony and font popularity, using fuzzy logic to predict aesthetic preferences. We collected our own dataset, consisting of nearly 200 popular and frequently used website designs, to ensure relevance and adaptability to the dynamic nature of web design trends. Dominant colors from website screenshots were extracted using k-means clustering. The findings aim to improve understanding of the relationship between aesthetics and usability in website design.

## 1. Introduction

In the present-day world, the number of Internet users is increasing rapidly. More than 5.35 billion people, representing over two-thirds of the global population, use the internet [1]. Websites are core to the Internet, providing a platform for sharing information and accessing online resources. They enable users to access various content and services, making the Internet more useful for everyone. However, the probability of retaining a user decreases as the number of websites increases. To differentiate yourself from the competition, it is important to focus on both the quality of content and the overall user experience.

First impressions are crucial in determining personality attribution in various contexts [2]. This is especially true for websites, as they strongly influence users' perceptions of appeal and usability [3], [4]. The visual design of a website can elicit positive emotions, enhance user satisfaction, and capture visitors' attention [5]. It is possible to make a reliable decision on whether users like the visual appeal in just 50 milliseconds [6]. Lindgaard et al.'s research illustrates how visual appeal may be closely linked to other design elements, such as layout and color, influencing overall impression.

Garrett [7] considers aesthetics a subjective judgment, as it varies from person to person and is highly dependent on that person's culture and personality. Some researchers have created models for measuring aesthetic values when considering it as a measurable object. If we view aesthetics as something that can be measured, then models for assessing aesthetic values can be developed, similar to the work of Zane et al. [8] and Filonik and Baur [9]. However, these works do not consider all aspects of the site, and only a few people participated in assessing aesthetics.

The paper is structured as follows. Section I is this introduction. Sections II and III present the main provisions and literature review. Next, Section IV describes methods, including data collection and annotation and a fuzzy approach. Section V presents experimental results along with the performance evaluation and discussion. Finally, concluding remarks are given in Section VI.

## 2. Main Provisions

The current paper presents a model that utilizes the fuzzy logic approach to determine websites' aesthetic preferences using color harmony and font aesthetics. Over 200 sites were collected. The overall methodology is presented in Fig. 1.

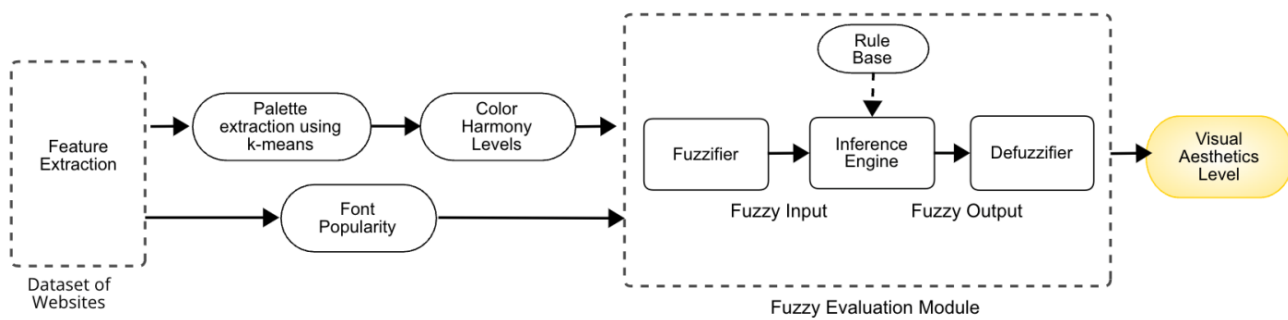

Figure 1. Methodology figure illustrating the proposed approach

The contributions of this study are:
• Application of Fuzzy Logic to create a model in website design aesthetics. We propose a new way to quantify subjective elements, such as color harmony and font aesthetics, and determine the site's aesthetics.
• We collected our dataset of the most popular and frequently used website designs from the Internet. This is important due to the dynamic nature of web design trends. Our analysis focuses on projects that gained prominence and were frequently used in 2024.

## 3. Literature Review

In this section, we provide an overview of measuring aesthetics.

Aesthetics can be defined as the philosophy of beauty and the perception of what is visually pleasing [10]. Lee and Koubek [11] suggest that the usability of content is processed later than perception. Several authors have pointed out [12], [13], and [14] that the first impression of a website is greatly influenced by its aesthetics. It is now known that our immediate emotional reaction to products, based on how they look, strongly affects how we later judge their usability [15]. Beautiful websites are seen as more usable [6], [16]. Several works applied a fuzzy approach to modeling aesthetic preferences [17], [18], [19].

Developers strive to create web pages that can create a specific experience and fit well into the overall website strategy [20]. In doing so, they carefully select various components such as text, photographs, and other media elements, complementing them using a variety of styles, color palettes, fonts, and animations. These elements work together to create a unique aesthetic and provide a cohesive visual impact to site visitors.

According to study [21], the use of color is crucial in the design of a website. It's important to consider the diverse range of users who will access the site, including those who may be color blind and unable to differentiate certain colors. Therefore, choosing colors should be carefully considered when developing a website. Additionally, "balance" is also significant. As mentioned in study [22], a balanced website makes users feel psychologically balanced.

Choosing the right distance between website elements such as photographs, color schemes, fonts, and others is also important. There is a close relationship between the number of design elements and the space they occupy, so it's crucial to maintain a balance [23], [24]. Designing a website with appropriate white space can enhance both its usability and aesthetics [24].

When evaluating the visual appeal of websites, it is important to consider the elements mentioned earlier and the design style, which is divided into 21 types by study [25].

The aesthetics of a website significantly impact its perception and usability. These include emotional reactions, color and balance in design, and spatial relationships between elements.

## 4. Materials and Methods

4.1. Data
4.1.1. Data Collection

For the experiments, a dataset containing approximately 200 websites was collected 1. The process of collection and preparation involved several key stages. The first phase focused on gathering all site links and their associated names. The selection of sites was based on frequent use and other significant criteria, ensuring their relevance and popularity among users. When selecting websites, particular attention was paid to including diverse fields to cover various topics and interests. The effort was made to incorporate sites in different languages to reflect the diversity of cultural contexts, thereby making the dataset more universal and beneficial for analysis. Subsequently, the process moved on to capturing complete screenshots of each site's homepage. This was achieved by utilizing a specific web browser extension in Google Chrome called "GoFullPage - Full Page Screen Capture."

Finally, the developer tools panel in the browser was utilized to ascertain the fonts used on each site. Particular attention was paid to the most frequently used font family, as this often represents the primary typographic choice of the website designers.

The result of the collected dataset can be seen in Table 1.

Table 1 – Sample records from dataset

| Website ID | Website Name | Website Link | Website Category | First color | Second color | Third color | Font-Family |
|---|---|---|---|---|---|---|---|
| 1 | YouTube | https://www.youtube.com/ | Social Media | #101010 | #505051 | #a0a0a0 | Roboto |
| 2 | Informburo | https://informburo.kz/ | News | #ffffff | #181818 | #e72116 | Roboto |
| 3 | Telegram | https://telegram.org/ | Messaging | #ffffff | #fdd626 | #1a1f27 | Lucida Grande |
| 4 | Naver Dictionary | https://dict.naver.com/ | Education | #1f1f1f | #6b6c6c | #aeaeae | Helvetica Neue |
| 5 | Naver | https://www.naver.com/ | Web Portal | #fefefe | #bcbdbc | #03c75b | Malgun Gothic |
| 6 | IEEE | https://ieeexplore.ieee.org/ | Academic | #ffffff | #18455a | #0294e1 | Helvetica |
| 7 | Kaspi.kz | https://kaspi.kz/ | Finance | #f5f5f | #0303 | #f044 | Roboto |

| Website ID | Website Name | Website Link | Website Category | First color | Second color | Third color | Font-Family |
|---|---|---|---|---|---|---|---|
| | | | | 5 | 03 | 34 | |
| 8 | Apple | https://www.apple.com/ | Technology | #f5f6f6 | #010101 | #424242 | SF Pro Text |
| 9 | Netflix | https://www.netflix.com/kz/ | Entertainment | #010101 | #fcfcfc | #434343 | Bebas Neue Font |
| 10 | KFC | https://www.kfc.kz/ | Food | #ffffff | #e3012b | #440f1a | Cera |
| ... | ... | ... | ... | ... | ... | ... | ... |
| 200 | HealthLine | https://www.healthline.com/ | Health | #f8f8f8 | #010101 | #fcba44 | Proxima Nova |

The dataset comprises information from more than 200 sites collected between March and May 2024. Following the data collection phase, the information can be organized and distributed based on categories, color schemes, and font types. These sites have been divided into categories, and the relationship between the distribution of these categories is depicted in Fig. 3.

Figure 2. Collected dataset

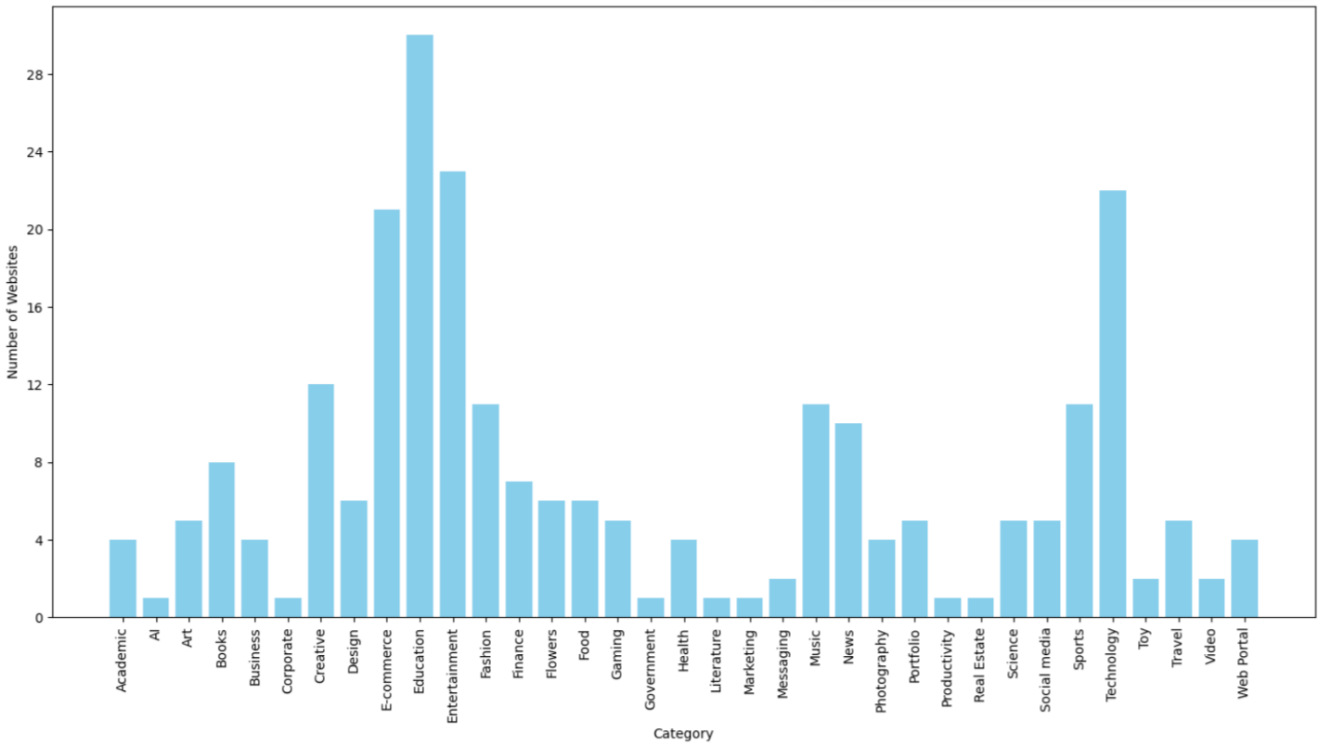

Figure 3. Distribution of collected websites by categories

Sample images of the website from our Dataset are presented in Fig. 4

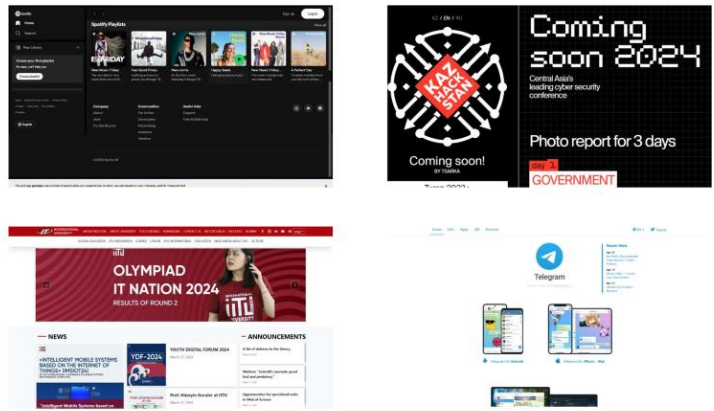

Figure 4. Sample images from our dataset. 1. Spotify. 2. KazHackStan. 3. IITU. 4. Telegram.

4.2. Fuzzy Sets and Logic

Fuzzy sets and logic provide a framework for dealing with uncertainty and imprecision, which are inherent in many real-world problems. Unlike classical sets, where an element belongs or does not belong to a set, fuzzy sets allow partial membership.

A fuzzy set A in a universe of discourse X is characterized by a membership function $\mu_A(x)$ that assigns to each element x in X a degree of membership in the interval [0,1] [26]. Mathematically, it is defined as:

$$\mu_A: X \to [0,1]$$

The basic operations in fuzzy logic include AND, OR, and NOT, which correspond to intersection, union, and complement in fuzzy set theory. For example, the fuzzy AND operation is defined as:

$$\mu_{A \cap B}(x) = \min(\mu_A(x), \mu_B(x))$$

A fuzzy inference system (FIS) is a framework for reasoning with fuzzy logic. It consists of the following steps [26]:
1. *Fuzzification.* Convert crisp inputs into fuzzy sets using membership functions.
2. *Rule Evaluation.* Apply fuzzy rules to the fuzzy inputs to obtain fuzzy outputs.
3. *Aggregation.* Combine the fuzzy outputs into a single fuzzy set.
4. *Defuzzification.* Convert the aggregated fuzzy set into a crisp output.

Fuzzy rules are an essential component of fuzzy logic systems. A fuzzy rule is generally expressed as an IF-THEN statement, where the IF part is the antecedent and the THEN part is the consequent. A fuzzy rule can be written as:

$$\text{IF } x_1 \text{ IS } A_1 \text{ AND } x_2 \text{ IS } A_2 \text{ THEN } y \text{ IS } B$$

where $x_1, x_2$ are input variables, $A_1, A_2$ are fuzzy sets representing the antecedents, and y is the output variable with fuzzy set B representing the consequent.

Fig. 5 depicts input fuzzy sets, Color Harmony and Font Popularity. Fig. 6 shows output fuzzy sets representing a website's level of Visual Aesthetics. Table 2 illustrates fuzzy rules used in the fuzzy inference system.

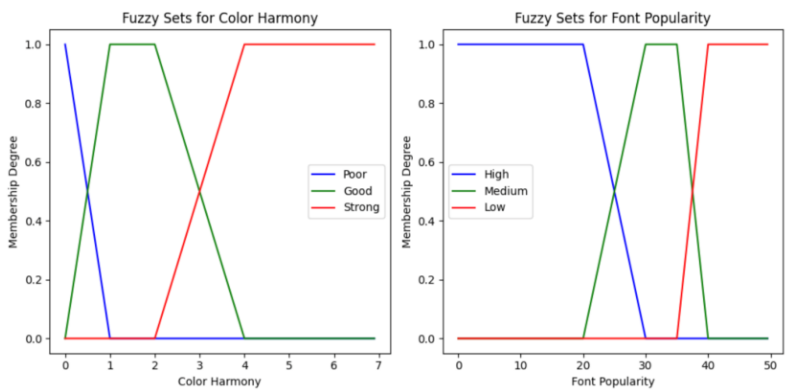

Figure 5. Input fuzzy sets, Color Harmony and Font Popularity

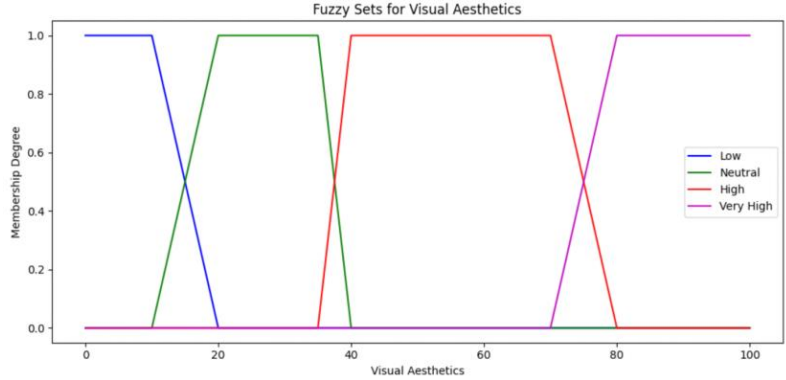

Figure 6. Output fuzzy sets representing the level of Visual Aesthetics in a website

Table 2 – Fuzzy rules in the knowledge base

| Rules | Color Harmony | Font Popularity | Visual Aesthetics |
|---|---|---|---|
| Rule 1 | Poor | Low | Low |
| Rule 2 | Poor | Medium | Low |
| Rule 3 | Poor | High | Neutral |
| Rule 4 | Good | Low | Neutral |
| Rule 5 | Good | Medium | High |
| Rule 6 | Good | High | High |
| Rule 7 | Strong | Low | Neutral |
| Rule 8 | Strong | Medium | Very High |
| Rule 9 | Strong | High | Very High |

4.3 K-means Clustering for Palette Extraction

K-means clustering is a popular unsupervised ML algorithm for partitioning a dataset into distinct, non-overlapping groups or clusters. It is widely used in various fields, including image processing, for tasks such as palette extraction, where the goal is to identify the dominant colors in an image.

The k-means clustering algorithm aims to partition n observations into k clusters, where each observation belongs to the cluster with the nearest mean. The algorithm consists of the following steps:
1. Randomly select $k$ initial cluster centroids from the dataset.
2. Assign each data point to the nearest cluster centroid based on the Euclidean distance. The assignment step can be represented as:

$$C_i = \{x_p : \| x_p - \mu_i \|^2 \leq \| x_p - \mu_j \|^2 \ \forall j,\ 1 \leq j \leq k\}$$

where $x_p$ is a data point, $\mu_i$ is the centroid of cluster $i$, and $C_i$ is the set of points assigned to cluster $i$.
3. Calculate the new centroids as the mean of all points assigned to each cluster. The update step can be represented as:

$$\mu_i = \frac{1}{|C_i|} \sum_{x_p \in C_i} x_p$$

where $|C_i|$ is the number of points in cluster $i$.
4. Repeat the assignment and update steps until the centroids converge, meaning the centroids no longer change significantly.

Euclidean Distance is defined as:

$$\| x_p - \mu_i \|^2 = \sum_{j=1}^{d} (x_{pj} - \mu_{ij})^2$$

where $x_{pj}$ is the j-th coordinate of data point $x_p$ and $\mu_{ij}$ is the j-th coordinate of the centroid $\mu_i$.

The objective of k-means is to minimize the total within-cluster variance, which is the sum of squared distances between points and their respective cluster centroids. The objective function can be represented as:

$$J = \sum_{i=1}^{k} \sum_{x_p \in C_i} \| x_p - \mu_i \|^2$$

As defined previously, the new centroid $\mu_i$ is the mean of all points $x_p$ in cluster $C_i$:

$$\mu_i = \frac{1}{|C_i|} \sum_{x_p \in C_i} x_p$$

In image processing, k-means clustering can extract the dominant colors in an image. Various color spaces can be employed for this purpose . The process involves treating each pixel's color value as a data point and clustering these points in the color space. The centroids of the clusters represent the dominant colors in the image, which can be used to create a color palette.

The steps for palette extraction are as follows:
1. Convert the image into a dataset of color values.

2. Apply the k-means clustering algorithm to this dataset.
3. The centroids of the clusters are the colors of the palette.

4.4. Color Wheel Harmony

We calculate color wheel harmony by extracting the dominant colors from a website using the K-means clustering algorithm. These dominant colors are then converted from RGB to HSV format, focusing particularly on the Hue component. The color wheel is divided into 12 segments, each corresponding to a primary, secondary, or tertiary color, and each segment represents a 30-degree slice of the wheel. By determining the Hue values of the dominant colors, we can locate their positions on this color wheel.

We then analyze these positions to identify the type of color harmony present. We used traditional harmonies, like "Monochromatic", "Complementary", "Split Complementary", "Triad", "Square", "Rectangular", "Analogous". For that, we used color-harmony 1.0.1 python package. For example, Complementary harmony is identified when colors directly oppose each other on the wheel, and Triad harmony consists of three colors evenly spaced around the wheel. Any combination that does not fit these categories is classified as "Other."

## 5. Results and Discussion

Color harmonies were extracted using the approach discussed in the previous section. Information about font popularity was parsed from the website www.myfonts.com. Now, we can illustrate the proposed approach with a specific example.

Fig. 7 illustrates the example of the resulting output, including dominant RGB colors, color wheel harmonies, and the palette illustration.

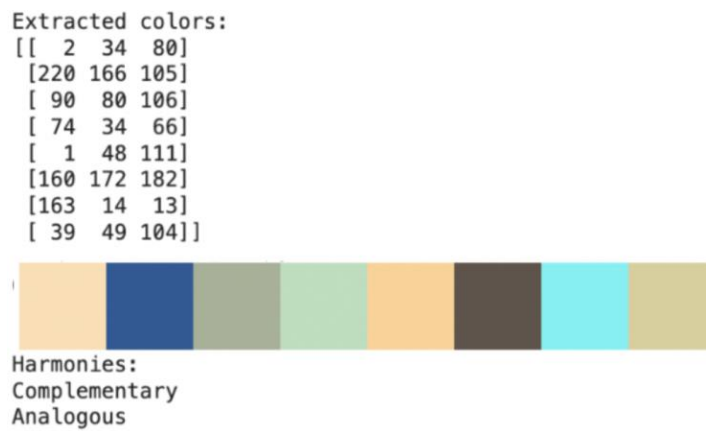

Figure 7. Example of extracted dominant RGB colors, wheel harmonies and palette illustration

To simulate the fuzzy system, we must first specify the inputs and then use the defuzzification approach. Consider the following example: the inputs are Color Harmony and Font Popularity with values of 3 and 37, respectively. Fuzzy aggregation then combines the output membership functions using the maximum operator. Next, we defuzzify to obtain a final result, which is done using the centroid approach. Aggregation using fuzzy criteria yielded an overall Visual Aesthetics score of 57.9%. This process effectively demonstrates how the fuzzy system simulates and visualizes the results.

```python
# Define fuzzy rules
rule1 = ctrl.Rule(color_harmony['Poor'] & font_popularity['Low'], visual_aesthetics['Low'])
rule2 = ctrl.Rule(color_harmony['Poor'] & font_popularity['Medium'], visual_aesthetics['Low'])
rule3 = ctrl.Rule(color_harmony['Poor'] & font_popularity['High'], visual_aesthetics['Neutral'])
rule4 = ctrl.Rule(color_harmony['Good'] & font_popularity['Low'], visual_aesthetics['Neutral'])
rule5 = ctrl.Rule(color_harmony['Good'] & font_popularity['Medium'], visual_aesthetics['High'])
rule6 = ctrl.Rule(color_harmony['Good'] & font_popularity['High'], visual_aesthetics['High'])
rule7 = ctrl.Rule(color_harmony['Strong'] & font_popularity['Low'], visual_aesthetics['Neutral'])
rule8 = ctrl.Rule(color_harmony['Strong'] & font_popularity['Medium'], visual_aesthetics['Very High'])
rule9 = ctrl.Rule(color_harmony['Strong'] & font_popularity['High'], visual_aesthetics['Very High'])
```

Figure 8. Defining rules using scikit-fuzzy

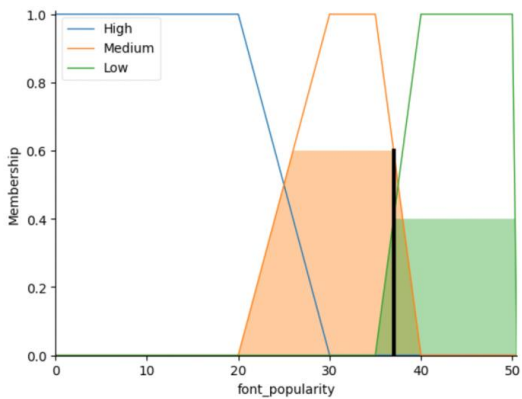

a. Applying input 3 Color Harmony count fuzzy set

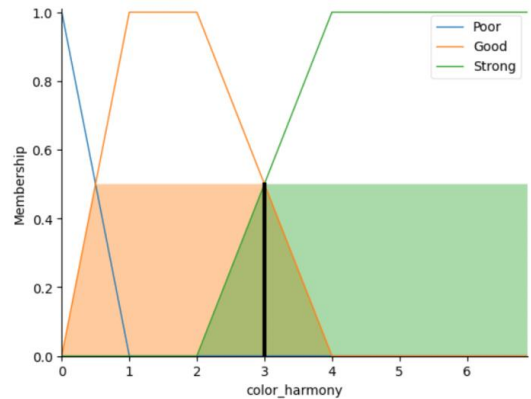

b. Applying input 37 on Font popularity fuzzy set

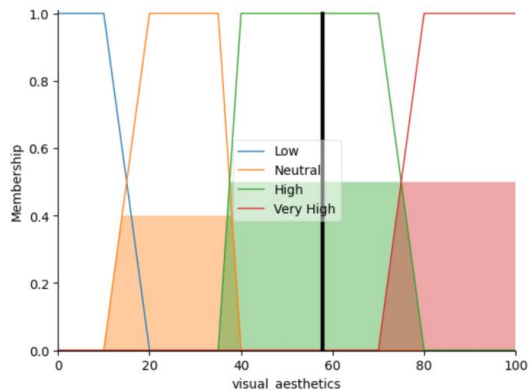

c. Aggregated Membership and Result, 57.9%

Figure 9. Simulation Results.

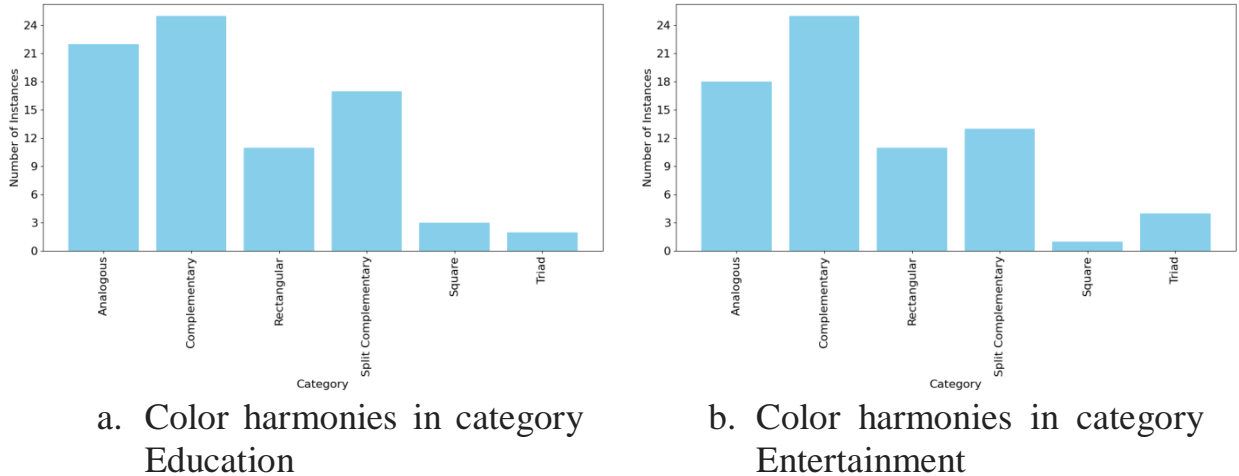

a. Color harmonies in category Education

b. Color harmonies in category Entertainment

Figure 10. The distribution of color harmonies.

The distribution of color harmonies in categories Education and Entertainment is presented in Fig. 10. As we see, color harmonies do not vary significantly across contexts, which partially supports previous research findings [19].

## 6. Conclusion

This study utilizes fuzzy logic and ML to develop a model predicting website aesthetics. The design of a website is critical because it significantly influences the user's first impression, often formed within milliseconds. A website's visual appeal can strongly affect how users judge its ease of use. A well-designed website is perceived as more trustworthy and usable. Therefore, considering aesthetics in website design enhances user experience and usability.

Our findings can empower developers and designers to design effective and visually captivating websites for users.

As for limitations, we consider only one website page, neglecting others. So, animations were not taken into account. Some websites were simple but had image content, which is part of information, not design. We also plan to integrate emotion-aware aesthetic assessment, considering, for example, Russell's circumplex model of affect, as in similar studies related to music [28].

## 7. Information on funding

This research has been funded by the Science Committee of the Ministry of Science and Higher Education of the Republic of Kazakhstan (Grant No. AP22786412)